\documentstyle[pra,aps,preprint,epsf]{revtex}

\begin{document}
\draft
\title{Experimental Study of A Photon as A Subsystem of An Entangled
Two-Photon State}
\author{Dmitry V. Strekalov\thanks{Present address: Department of Physics,
New York University, NY 10003},Yoon-Ho Kim, and Yanhua Shih}
\address{Department of Physics, University of Maryland, Baltimore County,\\
Baltimore, MD 21250}
\maketitle

\begin{abstract}
The state of the signal-idler photon pair of spontaneous parametric down
conversion is a typical nonlocal entangled pure state with zero entropy. The
precise correlation of the subsystems is completely described by the state.
However, it is an experimental choice to study only one subsystem and to
ignore the other. What can we learn about the measured subsystem and the
remaining parts? Results of this kind of measurements look peculiar. The
experiment confirms that the two subsystems are both in mixed states with
entropy greater than zero. One can only obtain statistical knowledge of the
subsystems in this kind of measurement.
\end{abstract}


One of the most surprising consequences of quantum mechanics is the
entanglement of two or more distant particles. The first example of a
two-particle entangled state was suggested by Einstein, Podolsky, and Rosen
in their famous {\em gedankenexperiment} in 1935 \cite{epr}. The EPR state
is a pure state of two spatially separated particles which can be written as,

\begin{equation}
\left| \Psi \right\rangle =\sum_{a,b}\delta \left( a+b-c_{0}\right) \left|
a\right\rangle \left| b\right\rangle  \label{eprst}
\end{equation}
where $a$ and $b$ are the momentum or the position of particle 1 and 2
respectively and $c_{0}$ is a constant. It is clear that state (\ref{eprst})
is a two-particle state; however, it cannot be factored into a product of
the state of particle 1 and the state of particle 2. This type of states was
defined by Schr$\ddot{o}$dinger as {\em entangled states} \cite{schro}.

One, perhaps the most easily accessible, example of an entangled state is
the state of a photon pair emitted in Spontaneous Parametric Down Conversion
(SPDC). SPDC is a nonlinear optical process from which a pair of
signal-idler photon is generated when a pump laser beam is incident onto an
 nonlinear optical crystal. The signal-idler two-photon state can be
calculated by first order perturbation from the SPDC nonlinear interaction
Harmiltonion \cite{Klyshkobook},

\begin{equation}
\left| \Psi \right\rangle =\sum_{s,i}\delta \left( \omega _{s}+\omega
_{i}-\omega _{p}\right) \delta \left( {\bf k}_{s}+{\bf k}_{i}-{\bf k}%
_{p}\right) a_{s}^{\dagger }(\omega ({\bf k}_{s}))\ a_{i}^{\dagger }(\omega (%
{\bf k}_{i}))\mid 0\rangle   \label{state}
\end{equation}
where $\omega _{j}$, {\bf k$_{j}$ (}j = s, i, p) are the frequency and
wavevectors of the signal (s), idler (i), and pump (p) respectively, $\omega
_{p}$ and {\bf k}$_{p}$ can be considered as constants, usually a single
mode laser is used for pump, $a_{s}^{\dagger }$ and $a_{i}^{\dagger }$ are
the respective creation operators for signal and idler photon. The delta
functions of the state ensure energy and momentum conservation. It is indeed
the conservation laws that determine the values of an observable for the
pair. Quantum mechanically, state (\ref{state}) only provides precise
momentum (energy) {\em correlation} of the {\em pair} but no precise
momentum (energy) determination for the signal photon and the idler photon.
In EPR's language: the momentum (energy) of neither the signal nor the idler
is determined by the state; however, if one is known to be at a certain
value the other one is determined with certainty. Notice also that state (%
\ref{state}) is a pure state. It provides a complete description of the
entangled two-photon system.

Following the creation of the pair, the signal and idler may propagate to
different directions and be separated by a considerably large distance.
If it is a free propagation, the state will remain unchanged except for the
gain of a phase, so that the precise momentum (energy) {\em correlation} of
the {\em pair} still holds. The conservation laws guarantee the precise
value of an observable with respect to the pair (not to the individual
subsystems). It is in this sense, we say that the entangled two-photon state
of SPDC is {\em nonlocal}. Quantum theory does allow a complete description
of the precise {\em correlation} for the spatially separated subsystems, but
no complete description for the physical reality of the subsystems defined
by EPR. It is in this sense, we say that quantum mechanical description
(theory) of the entangled system is{\em \ nonlocal}.

So far, our discussion involves no measurement.

In a type of measurements when ``joint detections'' are involved, for
example a coincidence detection for the SPDC pair, it corresponds to the
intensity correlation, $\left\langle \Psi \right| \hat{I}_{1}\otimes \hat{I}%
_{2}\left| \Psi \right\rangle $ or the fourth order correlation of the
fields, $\left\langle \Psi \right| \hat{E}_{1}^{(-)}\hat{E}_{2}^{(-)}\hat{E}%
_{2}^{(+)}\hat{E}_{1}^{(+)}\left| \Psi \right\rangle $. One may cooperate
spin (polarization for photon) correlations into the coincidence joint
detection too, as in the measurements for the EPR-Bohm state \cite{Bohm}. If
the correlations of the pair have been built up in the entangled
two-particle state from the beginning, it comes as no surprise that the
intensity correlation reflects {\em perfect} correlation (EPR, EPR-Bohm or
EPR-Bell type correlation) of the pair. The distance between the detectors
would not matter. There is no{\em \ action-at-a-distance} involved, if we
are willing to give up the classical EPR reality. The entangled state indeed
indicates and represents a very different physical reality: Does the signal
photon or idler photon have a defined momentum (energy) in state (\ref{state}%
)? No! Does the pair have a defined {\em total momentum} ({\em energy}) in
state (\ref{state})? Yes! In state (\ref{state}) the precise value of an
observable is determined in the form of {\em total} {\em value} by
conservation laws. In addition, one cannot ``assume'' or ``imagine'' two
individual wavepackets each associated with the signal photon and the idler
photon. It is a non-factorizable
two-dimensional ``wavepacket'' associated with the entangled two-particle
system \cite{Rubin}. For this very reason we have named the signal-idler pair the ``{\em %
biphoton''}. Many interesting phenomena involving biphoton have been
demonstrated in two-photon interferometry and in two-photon correlation type
experiments \cite{sample}. Several recent experiments have clearly shown
that the two-photon interference is not the interference between two
photons. It is not the signal and idler photon wavepackets but the
two-dimensional biphoton wavepacket that plays the role \cite{2and1}.

Recently, there has been a lot of interest in a type of measurements in
which only one subsystem of an entangled multi-particle state is measured
and the remaining parts are left undisturbed. One of the popular
misconceptions is to believe that the ``state'' of the undisturbed remaining
parts is completely determined by this kind of ``distance measurement'': if
the measurement of the subsystem either yields the result of a value for an
observable or indicates the ``state'' of that subsystem, the undisturbed
remaining parts is then ``forced'' into a ``pure state''. Do we have to
accept {\em action-at-a-distance} in quantum theory?

It is a fact that the experimentalist can choose to look at one part of the
entangled system and to ignore the other. The subsystems may well be
separated spatially. For instance, one can use a photon counting detector to
register a ``click'' event of the signal photon of the two-photon state of
SPDC (not a ``click-click'' event!) to study only the properties of the
signal and leave the idler undisturbed or to be predicted. What can we
predict for the idler photon in this kind of measurement? Before answering
this question, it may be better to ask first: ``what can we learn about the
signal photon in this kind of measurement?''

An interesting situation arises: while the two-photon state of SPDC is a
pure state, the respective states of the signal and idler photon are not.
The states of the individual signal and idler are both in thermal (mixed)
states. This has been pointed out by several researchers, e.g. \cite
{takahashi75,yurke87,klyshko96,cerf97}, from different perspectives.
Significance of two parts in a mixed state constitute a quantum mechanical
system in a pure state was emphasized by B. Yurke and M. Potasek \cite
{yurke87} as an example of purely quantum thermalization, that is obtaining
mixed states out of pure states in a Hamiltonian system. N.J. Cerf and C.
Adami \cite{cerf97} introduced the {\em mutual} ($S_{A:B}=S_{B:A}$) and {\em %
conditional} ($S_{A|B}$, $S_{B|A}$) entropy (or information) for a
two-particle system similar to the mutual and conditional entropies as
defined in classical probability theory:
\begin{equation}
S=S_{A|B}+S_{B|A}+S_{A:B},\quad S_{A}=S_{A|B}+S_{A:B},\quad
S_{B}=S_{B|A}+S_{A:B}.  \label{smisc}
\end{equation}
For an entangled two-particle system in a pure state (so that $S=0$), the
relations in (\ref{smisc}) give,
\begin{equation}
S_{A}+S_{B|A}=0,\qquad S_{B}+S_{A|B}=0.  \label{s0}
\end{equation}
The paradox of the whole system entropy $S$ being zero while an entropy of
its either part $S_{A}$ and $S_{B}$ are both positive (which is a formal
expression of the statement that the information contained in the whole
system is less than the information contained in its parts) is suggested
resolvable by letting the conditional entropy to take on negative values.

In this paper we report an experimental work along the lines of this
discussion. The reported experiment hinges on a typical Fourier spectroscopy
measurement. The schematic setup is shown in Fig. \ref{fig:setup}. The
measurement is based on a ``click'' type single photon detection; however,
the photon source is an entangled two-photon source of SPDC: a $3mm$ BBO ($%
\beta -BaB_{2}O_{4}$) crystal pumped by a $351.1nm$ CW Argon ion laser line.
The orthogonally polarized signal-idler photon pairs are generated by
satisfying the collinear degenerate (centered at wavelength $702.2nm$)
type-II phase matching condition \cite{Yariv}. The idler (extraordinary ray
of BBO) is removed by a polarizing beamspliter PBS. The signal (ordinary ray
of BBO) is then sent to a Michelson interferometer. A photon counting
detector is coupled to the output port of the interferometer through a $25mm$
focal lens. A $702.2nm$ spectral filter with Gaussian transmittance function
(bandwidth $83nm$ FWHM) is placed in front of the detector. The counting
rate of the detector is recorded as a function of the optical arm length
difference, $\Delta L$, of the Michelson interferometer.

The SPDC spectrum closely resembles a rainbow ranging from red to blue. The
spectrum collected by the $25mm$ focal lens is much wider then $83nm$. It
would be reasonable to expect a Gaussian spectrum with $83nm$ FWHM
(determined by the spectrum filter used for the detector) from the above
measurement. (This conjecture is different from the wrong belief or
imagination that each individual of the signal and idler photons is
associated with a Gaussian wavepacket.) On the contrary, we instead observed
an ``unexpected'' result, the observed spectrum is not Gaussian and its
width is only $2.2nm$ (far from $83nm$). The experimental data is reported
in Fig. \ref{fig:wpacket}. The envelope of the sinusoidal modulations (in segments)
is fitted very well by two ``notch'' functions (upper and lower part of the
envelope). The width of the triangular base is about $225\mu m$ which
corresponds to roughly a spectral band width of $2.2nm$.

To seek an explaination of this result, we must first examine the two-photon
state of SPDC. We cannot assume a state for either the signal photon or
idler photon. The single photon state is obtained by taking a partial trace
of the two-photon state density operator, integrating over the spectrum of
the idler and vice versa:
\begin{equation}
\hat{\rho}_{s}=tr_{i}\ \hat{\rho},\quad \hat{\rho}_{i}=tr_{s}\ \hat{\rho}
\label{ro}
\end{equation}
with
\begin{equation}
\hat{\rho}\equiv \left| \Psi \right\rangle \left\langle \Psi \right| ,
\label{stateII}
\end{equation}
where $\hat{\rho}$ the density matrix operator and $\left| \Psi
\right\rangle $ the two-photon state (\ref{state}).

First, it is very interesting to find that even though the two-photon EPR
state of SPDC is a pure state, i.e., $\hat{\rho}^{2}=\hat{\rho},$ the
corresponding single photon state of the signal and idler are not, i.e., $%
\hat{\rho}_{s,i}^{2}\neq \hat{\rho}_{s,i}$. This accords with the earlier
mentioned fact that the entropy of the system is zero (pure state) while
each subsystem has an entropy greater than zero (mixed state). The zero
entropy condition for a system in a pure state reflects the fact that the
quantum state $|\Psi \rangle $ provides a {\em complete} description of the
system . On the other hand, the mixed state of each subsystem only reveals
their statistical nature.

In the experiment, we realize a collinear degenerate type-II phase matching
\cite{Yariv}. This means that the SPDC crystal orientation is such that the
orthogonally polarized signal-idler pair with degenerate frequency $\omega
=\omega _{p}/2$, are emitted collinearlly. We select this direction by a set
of pinholes during the experimental alignment process. Then the integral in
Eq.(\ref{state}) can be simplified to an integral over a frequency detuning
parameter $\nu $, (the detailed calculation can be find in ref.\cite{Rubin}%
):
\begin{equation}
|\Psi \rangle =A_{0}\int d\nu \ \Phi (DL\nu )\ a_{s}^{\dagger }(\omega +\nu
)a_{i}^{\dagger }(\omega -\nu )\ |0\rangle .  \label{psinu2}
\end{equation}
where the $sinc$-like function $\Phi (LD\nu )$ followed from Eq.(\ref{state}%
) considering a finite length of the SPDC crystal \cite{Klyshkobook}. It
represents a spectral width of the two-photon state,
\begin{equation}
\Phi (DL\nu )=\frac{1-e^{-iDL\nu }}{iDL\nu },  \label{sinc}
\end{equation}
which is determined by the finite crystal length $L$ and, specifically for
the collinear degenerate type-II SPDC, by the difference of inverse group
velocities for the signal (ordinary ray) and the idler (extraordinary ray): $%
D\equiv 1/u_{o}-1/u_{e}$.

The constant $A_{0}$ is found from the normalization condition $tr\rho
=\langle \Psi |\Psi \rangle =1$ (dimensionless):
\[
A_{0}=\sqrt{\frac{DL}{4\pi }}.
\]

Substituting $|\Psi \rangle $ in the form of Eq.(\ref{psinu2}) into Eq.(\ref
{ro}) the density matrix of signal is calculated to be,
\begin{equation}
\hat{\rho}_{s}=A_{0}^{2}\int d\nu \ \left| \Phi (\nu )\right| ^{2}\
a_{s}^{\dagger }(\omega +\nu )\left| 0\right\rangle \left\langle 0\right| \
a_{s}(\omega +\nu )  \label{densityIs}
\end{equation}
where
\begin{equation}
\left| \Phi (\nu )\right| ^{2}={\rm sinc}^{2}\frac{DL\nu }{2}
\label{sincsignal}
\end{equation}

In (\ref{densityIs}) we consider a multimode (a continues frequency
spectrum) entangled system with a single quantum, $n=1$. The operator (\ref
{densityIs}) describes the statistical distribution of this quantum. This is
a good approximation since the coupling in SPDC is week and greater number
states $n>1$ that correspond to higher perturbation orders are extremely
unlikely. On the other hand, $n=0$ represents vacuum fields that do not
result in detections \cite{Squeezed}.

By now, we can understand very well the experimental results. (1) For a
spectrum of $sinc$-square function we do expect a double ``notch'' envelope
in the measurement and the base of the triangle, which is determined by
$DL$, is calculated to be $225\mu m$ (we have considered the optical
path difference is twice of the arm difference in Michelson interferometer),
corresponding to a $2.2nm$ bandwidth. The experimental result, from fitting,
is about $225nm$, which agrees well with the prediction \cite{Double}. (2)
We see that the spectrum of the {\em signal} photon is dependent on the
group velocity of the {\em idler} photon which is not measured at all in our
experiment. However, this comes as no surprise, because the state of the
signal photon is calculated from the two-photon state by integrating over
the idler modes. (3) We also see immediately that $\hat{\rho}_{s}^{2}\neq
\hat{\rho}_{s}$, so the signal and idler single-photon states are both mixed
states. It is then straightforward to evaluate numerically the Von Neuman
entropy $S$ \cite{inf} of the signal (or idler) subsystem,
\begin{equation}
S_{s}=-tr[\hat{\rho}_{s}\log \hat{\rho}_{s}],  \label{nentropy}
\end{equation}
based on the ``double notch'' fitting function. Note that operator (\ref
{densityIs}) is diagonal. Taking its trace is simply to perform an
integration over the frequency spectrum with the spectral density of Eq.(\ref
{sincsignal}). To compute the integral of Eq.(\ref{nentropy}) for the
density matrix $\hat{\rho}_{s}$ of Eq.(\ref{ro}), we replace variable $\nu $
by a dimensionless variable $DL\nu /2$ and evaluate the integral
numerically. The calculation yields,
\[
S_{s}\approx 6.4>0.
\]
This again indicates the statistical mixture nature of the state of a photon
(subsystem) in an entangled two-photon system.

Based on the experimental data, we conclude that the entropy of signal and
idler are both greater then zero (mixed state); while the entropy of the
signal-idler two-photon system is zero (pure state). This may mean that
negative entropy is present somewhere in the system, perhaps in the form of
the conditional entropy \cite{cerf97}. By definition of the conditional
entropy, one is tempted to say that {\em given the result of a measurement
over one particle, the result of measurement over the other must yield
negative information.} This paradoxical statement is similar to and in fact
closely related to the EPR ``paradox''. We suggest that the paradox comes
from the same philosophy.

Conclusion: In these kind of measurements, in which the experiment only
measures a subsystem of an entangled multi-particle system and leave the
remaining parts undisturbed, one can only obtain statistical knowledge of
the subsystems. Neither the measured subsystem nor the remaining parts is in
pure state. The individual subsystems are described statistically by the
quantum theory before the measurement and after the measurement. The
measurements can never ``force'' the undisturbed subsystems into a pure
state. Again, we emphasize that no {\em action-at-a-distance} in any format.

We gratefully acknowledge the many useful discussions with M.H. Rubin. This
work was supported, in part, by the U.S. Office of Naval Research and
National Security Agency.

\newpage

\begin{figure}[tbp]
\caption{ Schematic of the experimental set up. A Michelson interferometer
is used to study the spectrum of the signal of SPDC. The SPDC spectrum
closely resembles a rainbow ranging from red to blue. A band pass spectral
filter centered at $702.2nm$ with $83nm$ FWHM of a Gaussian transmittance
function, is placed in front of the photon counting detector.}
\label{fig:setup}
\end{figure}

\begin{figure}[tbp]
\caption{Experimental data indicated a ``double notch'' envelope of the
interference pattern. The X-axis, $\Delta L$ in $\mu m$, is the optical
arm
difference of the Michelson interferometer. Each of the doted single
vertical segment contains many cycles of sinusoidal modulations. The spike
at $\Delta L=0$, usually called ``white light condition'' for
observing``white light'' interference, is a broad band interference pattern
which is determined by the spectral filter. Two vertical lines show $%
+122.5\mu m$ and $-122.5 \mu m$.}
\label{fig:wpacket}
\end{figure}

\newpage \centerline{\epsffile{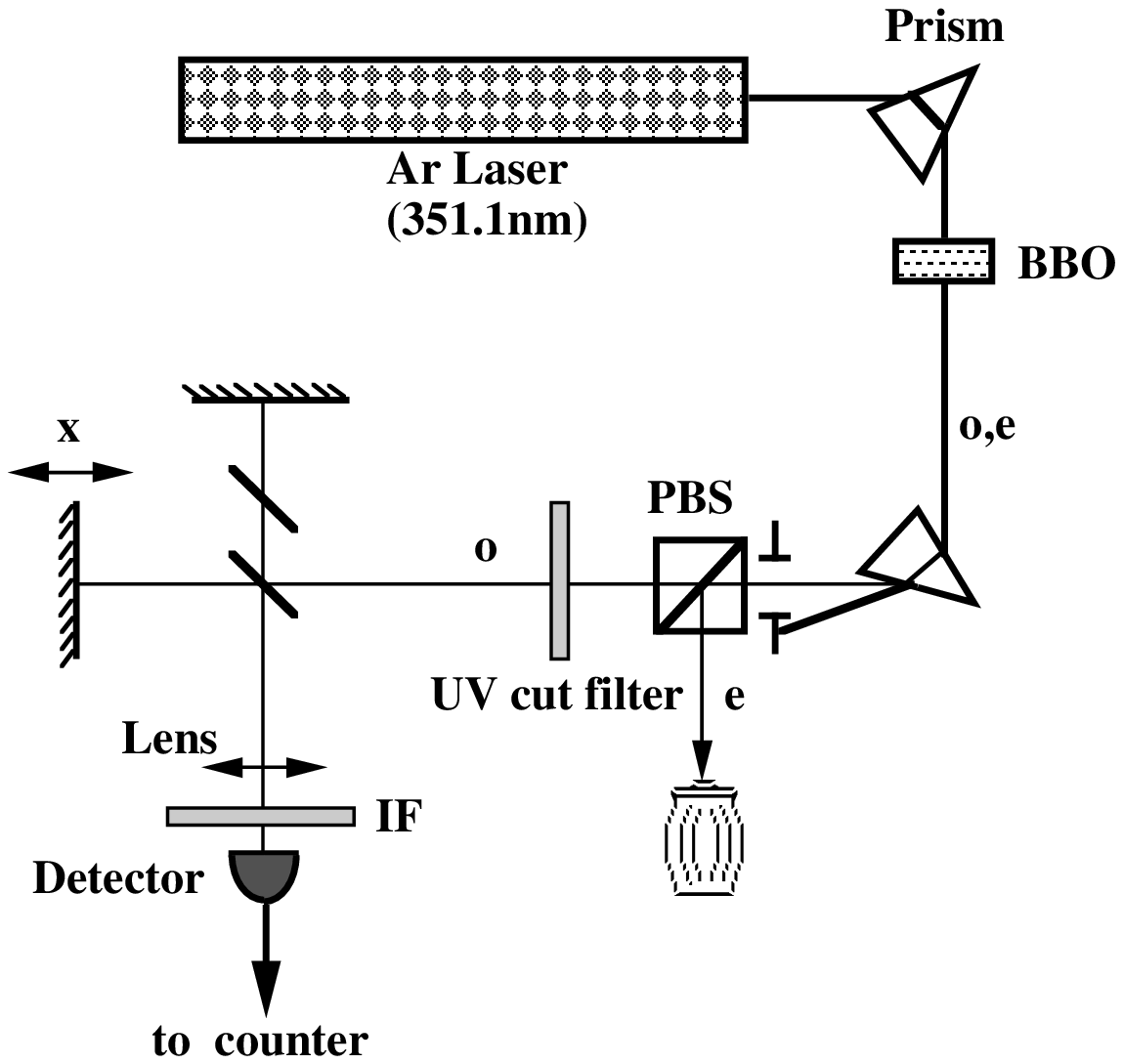}} \vspace{1cm} Figure 1. Dmitry V.
Strekalov, Yoon-Ho Kim, and Yanhua Shih

\newpage \centerline{\epsffile{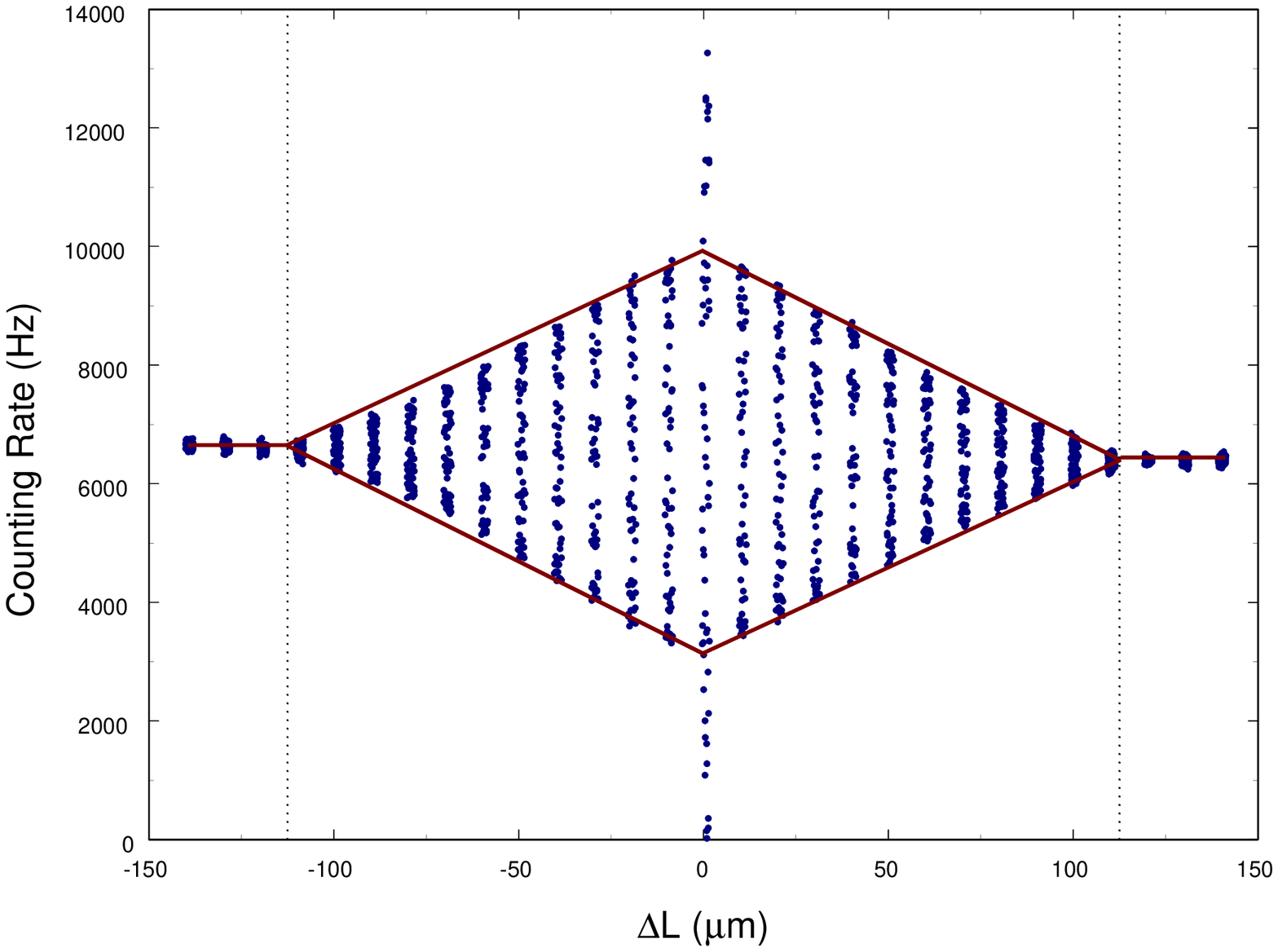}} \vspace{1cm} Figure 2. Dmitry
V. Strekalov, Yoon-Ho Kim, and Yanhua Shih


\begin{references}
\bibitem{Squeezed}  Our approach therefore is quite different from the early
``squeezed state'' studies, e.g. \cite{takahashi75,yurke87} where the signal
and idler radiation of SPDC is shown to be in the thermal (and hence mixed)
state based on a single-mode model which allow all number-states but omits
entanglement.
\end{references}
\end{document}